\documentclass{article}
\usepackage{spconf,amsmath,graphicx}

\usepackage{cite}
\usepackage{tabularx, makecell,array,booktabs, multirow, subfigure, amssymb, footnote, url}
\newcolumntype{P}[1]{>{\centering\arraybackslash}p{#1}}

\usepackage{xcolor}
\usepackage{color}

\usepackage{amssymb}
\usepackage{pifont}

\newcommand{\fref}[1]{Figure~\ref{#1}}
\newcommand{\tref}[1]{Table~\ref{#1}}
\newcommand{\sref}[1]{Section~\ref{#1}}

\title{Few-Shot Musical Source Separation}
\name{Yu Wang$^{\sharp}$\sthanks{This work was performed while interning at Spotify. This work is partially supported by National Science Foundation award 1544753 \& 1955357.}, Daniel Stoller$^{\flat}$, Rachel M. Bittner$^{\flat}$, Juan Pablo Bello$^{\sharp}$}
\address{$^{\sharp}$Music and Audio Research Laboratory, New York University, NY, USA\\
$^{\flat}$Spotify, New York, NY, USA}
\begin{document}
\ninept
\maketitle

\begin{abstract}
Deep learning-based approaches to musical source separation are often limited to the instrument classes that the models are trained on and do not generalize to separate unseen instruments. To address this, we propose a few-shot musical source separation paradigm. We condition a generic U-Net source separation model using few audio examples of the target instrument. 
We train a few-shot conditioning encoder jointly with the U-Net to encode the audio examples into a conditioning vector to configure the U-Net via feature-wise linear modulation (FiLM).
We evaluate the trained models on real musical recordings in the MUSDB18 and MedleyDB datasets. We show that our proposed few-shot conditioning paradigm outperforms the baseline one-hot instrument-class conditioned model for both seen and unseen instruments. To extend the scope of our approach to a wider variety of real-world scenarios, we also experiment with different conditioning example characteristics, including examples from different recordings, with multiple sources, or negative conditioning examples.


\end{abstract}


\section{Introduction}
\label{sec:intro}
Musical source separation (MSS) is a well-studied problem which has seen rapid progress in recent years~\cite{cano2018musical}, where the goal is typically to separate the sound of a particular instrument from a mixture recording.
Systems using various deep learning-based approaches~\cite{simpson2015deepkaraoke,uhlich2017improving,jansson2017singing,luo2017,stoller2018wave,takahashi2018,spleeter2020,takahashi2021} have achieved impressive results, in particular for singing voice separation.
However, these models are typically trained to separate one particular instrument class (e.g. vocals, drums).
In order to separate more than one instrument, more than one model is required.
More recent models have aimed to overcome this limitation and support the separation of various instruments using a single model via instrument class conditioning mechanisms~\cite{slizovskaia2019end,meseguer2019conditioned,seetharaman2019class,choi2021lasaft, samuel2020}.
However, these models are still limited to the instrument classes that the models were trained on and do not generalize to unseen instruments.

A few initial attempts at building deep learning-based query-by-example MSS have been made, where the model is designed or has an auxiliary benefit to perform separation based on an audio query~\cite{lee2019,kumar2018,seetharaman2019class,manilow2020}. However, these models have not been thoroughly evaluated on how they generalize to unseen classes. 
A similar recent effort has been made to develop a one-shot source separation model for general sounds~\cite{gfeller2021}.
While this system is evaluated on how well it separates seen and unseen sound classes given one audio example from the target, it has not been compared with any baseline, such as an existing class label-conditioning approach. 
Furthermore, it is evaluated on random mixtures of two sources, which are both less realistic and arguably much easier to separate than real musical recordings. 
A meta-learning approach was also recently proposed to achieve one-shot speech separation~\cite{wu2021one}. 
However, it has a very specific task definition of separating two speakers given one audio example per speaker, which is less applicable to general MSS. 
In addition, all of the systems mentioned above are relatively limited in their flexibility -- they consider using only a single query example which is isolated (single-sourced) and in some cases drawn from the same recording as the target, which may not always be available in real-world scenarios.



\begin{figure}
    \centering
    \includegraphics[trim=0 0.7cm 0 0.2cm, width=\columnwidth]{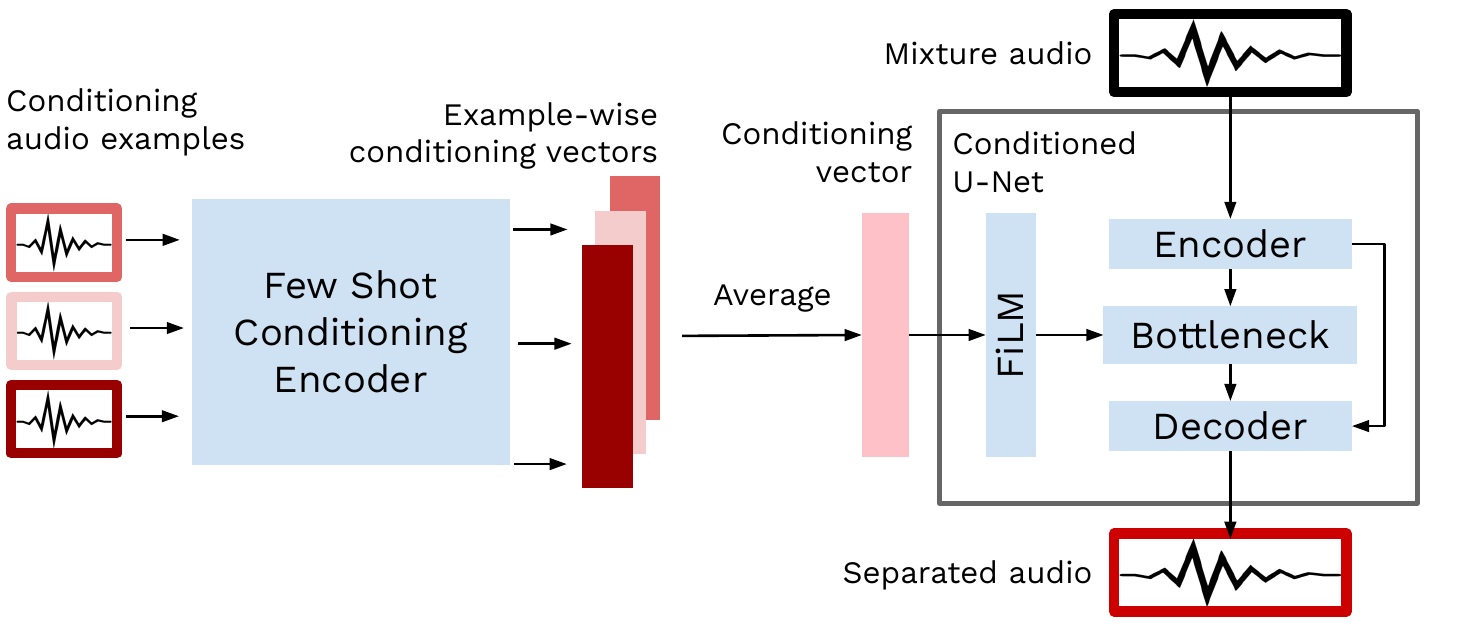}
    \caption{High-level illustration of the few-shot MSS paradigm.}
    \label{fig:model}
    \vspace{-1.7mm}
\end{figure}

In this work, we present a few-shot MSS paradigm illustrated in Figure~\ref{fig:model}. We condition a generic U-Net separation model using few audio examples of the target instrument.
We first embed the conditioning audio examples into a set of example-wise conditioning vectors using the few-shot conditioning encoder, and aggregate these vectors into a single conditioning vector by taking the average. 
We use Feature-wise Linear Modulation (FiLM)~\cite{perez2018} as the conditioning mechanism, which is inserted at the U-Net bottleneck layer. This mechanism allows us to incorporate side information, the conditioning vector, to configure the U-Net to separate different target instruments. We jointly train the few-shot conditioning encoder and the conditioned U-Net.    
Note that our proposed model is similar to~\cite{gfeller2021}, but we extend it from one-shot to few-shot with a simpler conditioning mechanism, where we apply the conditioning to the bottleneck layer only~\cite{slizovskaia2019end, meseguer2019conditioned}, instead of to multiple layers in both the encoder and decoder of the U-Net. 

We systematically compare our few-shot conditioned model against a class label-conditioned baseline approach. We quantitatively evaluate both approaches on seen and unseen instruments within real musical recordings. 
In addition, we explore how much the constraints on the conditioning scenarios in previous work~\cite{gfeller2021} can be relaxed. 
Besides extending from one-shot to few-shot, in particular, we consider providing (1) examples that are not drawn from the target recording, (2) examples where the desired target is not isolated, but is present within a mixture, and (3) negative examples of what \textit{not} to separate in addition to positive examples. 
We show that these relaxations result in systems which can be applied in a wider variety of applications. 
Audio examples from our few-shot MSS model are publicly available\footnote{\url{https://wangyu.github.io/few-shot-mss/}}.  
\section{Model Configurations}
\label{sec:models}

We base our separation model on the well-studied U-Net architecture~\cite{jansson2017singing,meseguer2019conditioned,cohen2019improving} in order to complement related work on conditioned source separation~\cite{meseguer2019conditioned,lee2019,slizovskaia2019end}.
While models with better separation performance exist~\cite{choi2021lasaft}, the purpose of this study is to explore the effects of different types of conditioning approaches and design choices, which can generally translate to better performance, rather than to strive for state-of-the-art results. 

\subsection{Instrument class-conditioned baseline}
We use a U-Net architecture very similar to that described in~\cite{meseguer2019conditioned} as the baseline.
The model is conditioned directly on instrument classes (via one-hot encoding) and can be viewed as a standard supervised learning baseline. Here, we briefly describe our variant of the model. 
For the U-Net architecture, each of the 6 encoder layers consists of $n_i$ 5x5 2-D convolutions with a 2x2 stride, followed by batch normalization and a Leaky ReLU activation, where $n_0=16$ and $n_i = 2n_{i-1}$ for each subsequent layer $i$. The decoder mirrors the encoder by replacing 2-D convolutions with 2-D transpose convolutions. 
Additional skip connections are added between layers at the same hierarchical level in the encoder and decoder.

A one-hot vector that directly encodes the target instrument class is used as the conditioning vector $z_{\textit{class}}$. 
A FiLM layer is inserted after the U-Net bottleneck layer to transform the bottleneck feature $x$ via $FiLM(x) = \gamma (z_{\textit{class}}) \cdot x + \beta (z_{\textit{class}})$, where $\gamma$ and $\beta$ are learned parameters that scale and shift $x$ based on the conditioning information $z_{\textit{class}}$. 
In our work, $z_{\textit{class}}$ is 18-dimensional to support a wide variety of instrument classes. 
The y-axis in \fref{fig:medleydb} shows 14 of these instrument classes, and the remaining 4 are \textit{mallets}, \textit{pipe organ}, \textit{bagpipes}, and \textit{whistling}. 

The model takes three seconds of mono audio with 22.05 kHz sample rate as input and first computes a spectrogram with a hop size of 256 and a FFT size of 1024. The spectrogram is log-compressed, and input into the U-Net encoder. The output of the decoder is treated as a two-channel complex mask, where a sigmoid activation is applied on the magnitude as in~\cite{choi2018phase}. The mask is applied to the compressed spectrogram via complex multiplication, followed by decompression. Finally, the Inverse STFT is computed to retrieve the separated audio signal. The loss is computed as the sum of the SDR Loss~\cite{venkataramani2017adaptive} in the time domain and mean absolute error on the magnitudes in the time-frequency domain.



\subsection{Few-shot conditioning}

In our proposed few-shot MSS paradigm, the external information is based on few audio examples of the target instrument instead of its class label.
We train a few-shot conditioning encoder jointly with the U-Net to generate a learned conditioning vector $z_{\textit{few-shot}}$ based on few target examples.

The conditioning encoder consists of four convolution blocks, each of which has a 64-filter 3x3 convolution, a batch normalization layer, a ReLU activation layer, and a 2x2 max-pooling layer. 
To allow our model to handle variable-length input, we apply max-pooling along the time dimension to the output of the convolution blocks. 
Finally, we flatten the feature map to output an embedding of size 512.
When multiple audio examples are used as conditioning (few-shot instead of one-shot), each example is used to generate a conditioning vector, and the example-wise vectors are averaged into a single conditioning vector.
We keep the conditioned U-Net architecture, FiLM conditioning mechanism, and loss terms the same as in the baseline model to systematically compare the instrument class conditioning and few-shot conditioning paradigms.

\section{Experimental Design}
To compare the proposed few-shot MSS with the baseline approach conditioned on instrument class, we first train the models using a combination of several datasets, and evaluate the trained models on real music datasets, MUSDB18~\cite{musdb18} and MedleyDB~\cite{bittner2014}. 
For the few-shot model, we further experiment with different evaluation setups with different conditioning example characteristics as well as including negative conditioning examples.  

\label{sec:exp}
\subsection{Training Datasets}
In order to support a wide variety of instrument classes, we train the models on a combination of several multitrack datasets consisting of solo-instrument stems that are linearly mixed to create full song recordings.
The first dataset, $D_{\textit{real}}$, is a private dataset containing 313 real-world multi-tracks, with an instrument bias towards pop/rock music meaning that vocals, drums, bass, guitars, and synthesizers dominate the distribution.
The second, $D_{\textit{synth}}$, is the public Slakh2100-redux~\cite{manilow2019} dataset, containing 1710 multi-tracks synthesized from MIDI with high-quality synthesizers. 
This dataset contains no vocals, and by its nature only contains synthetic versions of each instrument.
$D_{\textit{solo}}$ is a private dataset of 627 real solo instrument recordings, roughly evenly distributed across the 18 instrument classes.
These recordings are randomly combined to generate multi-tracks with up to 5 instruments.

\subsection{Training setup}
To generate a training example, we randomly sample a dataset $D$ from the set $\{D_{\textit{real}}, D_{\textit{synth}}, D_{\textit{solo}}\}$, a multi-track $M$ from $D$, a chunk from $M$ as the input mixture, and finally an instrument track from the mixture as the target output. 
To train the few-shot models, we additionally sample $n$ random chunks from the original track of the target instrument as the conditioning examples. 
Note that we make sure the random chunks do not overlap with the target output. 
We experiment with $n$ between one and five (1-shot to 5-shot) to examine how this choice correlates with model performance. 
All models are trained using a batch size of 16 and a learning rate of 0.001 with the Adam optimizer and early stopping.

\subsection{Evaluation setup}
We evaluate the trained MSS models on the 50 test tracks in the MUSDB18 dataset. For each track, we have its mixture and isolated tracks of \textit{vocals}, \textit{drums}, \textit{bass}, and \textit{other} in stereo format with a 44.1k Hz sample rate. 
An \textit{other} track includes all instruments in a mixture other than vocals, drums, and bass. 
It can vary greatly over time with different instruments present at different times, making conditioning using random examples potentially incoherent.
Therefore, we do not evaluate on separating \textit{other} in this work. 
Note that we did not include the training set of MUSDB18 into our training data. 
The training and testing data are therefore from different distributions, resulting in a more challenging setup compared to most of the previous works on MSS.

To evaluate a test track, we downsample the audio to 22 kHz and divide the mixture into 3-second input segments with 50\% overlap. 
We sample $n$ random chunks 
from the target instrument track as the conditioning examples for the few-shot models, and run inference on each 3-second chunk. We reconstruct the final signal using overlap-add, and upsample to 44.1 kHz. 
We compute the signal-to-distortion ratio (SDR)~\cite{vincent2006} per track using the \texttt{museval} package as our evaluation metric, and average the SDR across tracks.
We repeat this process 10 times for each track with different random conditioning examples when evaluating few-shot models, reporting the mean and standard deviation of means over each of the 10 test iterations.

While MUSDB18 is the most common MSS benchmarking dataset, it only contains isolated tracks for three instrument classes. 
Since our baseline model supports up to 18 instruments, we additionally evaluate the baseline and the best performing few-shot model on the MedleyDB dataset. 
We filter out multi-tracks with bleed, map all instruments to the predefined 18-instrument vocabulary, and generate sub-mixes in which we keep only one stem per instrument class. 
The resulting MedleyDB evaluation set contains 90 tracks with 14 instrument classes.


\begin{table}[t]
\centering
\footnotesize
\begin{tabular}{cccc}
\toprule
Method & Vocals & Drums & Bass \\
\midrule
Baseline & $\footnotesize 4.07$ & $4.35$ & $3.06$ \\
\midrule
1-shot & $4.23 \pm 0.88$ & $4.34 \pm 0.60$ & $2.71 \pm 0.86$ \\
2-shot & $4.64 \pm 0.38$ & $4.54 \pm 0.39$ & \boldmath$3.27 \pm 0.41$ \\
3-shot & \boldmath$ 4.80 \pm 0.32$ & \boldmath$4.71 \pm 0.38$ & $3.16 \pm 0.45$ \\
4-shot & $4.58 \pm 0.39$ & $4.61 \pm 0.32$ & $3.08 \pm 0.50$ \\
5-shot & $4.72 \pm 0.35$ & $4.66 \pm 0.25$ & $3.26 \pm 0.38$ \\
\bottomrule
\end{tabular}
\caption{Mean SDR (dB) on MUSDB18 for the baseline and few-shot conditioned models with different \textit{n}. The standard deviation of the means for few-shot models are computed over 10 test iterations.}
\label{tab:onehot-vs-fewshot}
\end{table}

\subsection{Conditioning example characteristics}
\label{subsec:conditioning}
In our basic setup of training and evaluating the few-shot model, the selected conditioning examples have two main constraints: (1) they are drawn from the target recording, (2) they are single-sourced audio examples that only contain the target instrument. 
This is a relatively ideal scenario when considering real-world applications, where isolated examples from the target recording may not be available. 
To see how a few-shot model generalizes to more challenging and practical scenarios, we perform a set of experiments where we evaluate the model with one constraint relaxed at a time. 

To relax the first constraint, we draw $n$ conditioning examples from $n$ different recordings of the same instrument class within the test data. 
To relax the second constraint, we allow some conditioning examples to contain one additional non-target instrument, where each example has a 0.5 probability of being multi-sourced. 
To get a multi-source example from a multitrack, we randomly choose a non-target instrument and mix it with the target before we sample an example chunk. 
Note that we make sure only one consistent instrument, the target, is present across all conditioning examples.    

\subsection{Negative conditioning examples}
\label{subsec:neg-cond}
Besides providing target instrument examples to specify the desired output, we further explore if specifying \textit{unwanted} instruments through the same conditioning mechanism can help improve the separation. 
To do so, we extend our basic few-shot model to leverage additional \textit{negative conditioning examples}. 
During both training and evaluation, we draw $n$ additional negative examples, 
each of which contains a random non-target instrument that also exists in the input mixture. 
All conditioning examples are first embedded via the same few-shot conditioning encoder. 
Then we average the example-wise conditioning vectors over positive and negative examples separately. 
We concatenate the aggregated positive and negative conditioning vectors along the feature dimension, and convert the concatenated vector to a final conditioning vector via a fully-connected layer with a ReLU activation. 
The sizes of the resulting conditioning vector and the conditioned U-Net model are the same as in our basic setup.  


\section{Results}
\label{sec:results}
\subsection{Instrument class conditioning vs. few-shot conditioning}\label{sec:baseline-results}



In \tref{tab:onehot-vs-fewshot}, we show the performance of the instrument class conditioned baseline and few-shot conditioned models with different numbers of conditioning examples $n$ (the number of shots). 

First, the 1-shot model, conditioned on only one audio example, does not show a significant advantage over the baseline. 
It achieves higher SDR on \textit{vocals}, but performs worse on \textit{bass}. 
While we can provide more direct and specific information about the target via audio examples for conditioning, one random example may not be sufficient to capture the variations of the target within the entire song.
On the other hand, as $n$ increases, the few-shot model begins to outperform the baseline across all instruments.
This shows the advantage of extending existing query-based or one-shot MSS to few-shot MSS. 
We can now provide more conditioning examples to better capture the instrument variations in pitch, timbre, and playing technique, to achieve better separation.
Next, we see a general trend that as $n$ increases, the performance of the few-shot model initially increases and then stays relatively stable, while the standard deviation continues to decrease. 
Since we aggregate example-wise conditioning vectors by taking an average, we can expect these diminishing returns from including more examples once enough representative examples are present.
On the other hand, conditioning on more audio examples can mitigate the randomness from sampling and leads to more stable results. 
Therefore, in the remainder of our experiments, we focus on a 5-shot model for more robust performance.      

\begin{figure}
    \centering
    \includegraphics[trim=0 1.1cm 0 0, width=\columnwidth]{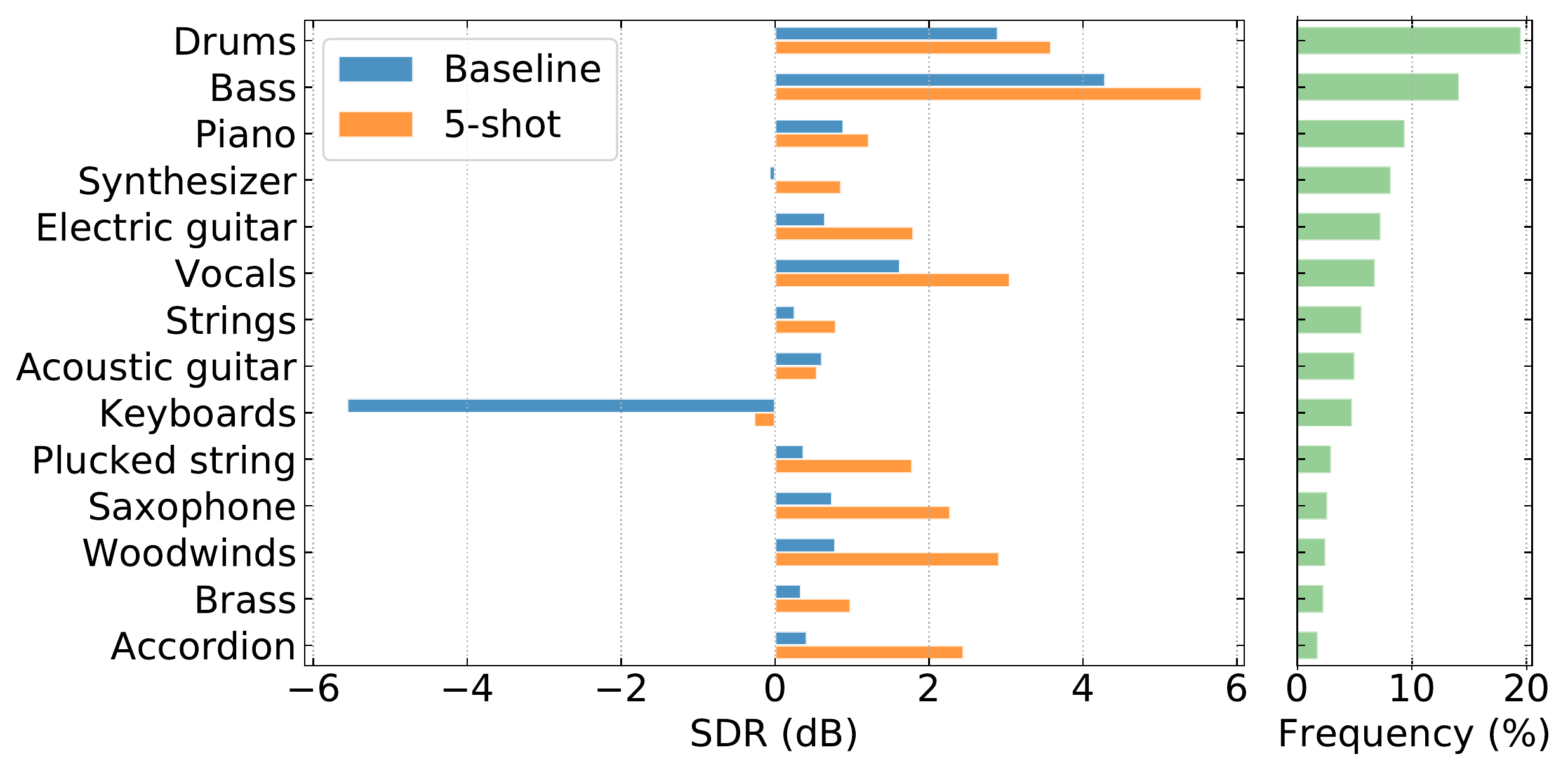}
    \caption{(Left) Median SDR on MedleyDB dataset for the baseline and 5-shot models. We report median SDR here due to high variance between tracks. (Right) Training data distribution.}
    \label{fig:medleydb}
\end{figure}

In addition to MUSDB18, we evaluate the baseline and 5-shot models on MedleyDB, which contains many more instrument classes. 
\fref{fig:medleydb} (left) first shows that the 5-shot model outperforms the baseline on 13 instruments out of the total of 14 with an overall SDR of 1.96 dB (0.56 dB for the baseline). 
Next, we sort the instruments 
by their occurrences in the training data based on the distribution shown in \fref{fig:medleydb} (right). 
We see that the 5-shot model can not only effectively separate instruments that are relatively rare in our training data, but also generally outperforms the baseline by larger margins on these rare classes, for example, \textit{Keyboards}, \textit{Woodwinds}, and \textit{Accordion}. This suggests that the advantage of few-shot MSS is more significant on rare instruments.
From these results, we see that few-shot MSS is applicable to many instruments across different datasets. In the remainder of our experiments, for brevity we focus on the MUSDB18 dataset.


\subsection{Separating unseen instruments}
One of the advantages of the few-shot MSS model is that it is able to handle new, unseen instruments at inference time without being limited by the training vocabulary. 
However, our training data has a large instrument vocabulary while MUSDB18 only contains three common instruments. 
To quantitatively evaluate how well models can separate unseen instruments, we remove all \textit{vocals}, \textit{drums}, and \textit{bass} target examples from our training data and retrain the baseline and few-shot models with the updated training set. This ensures that the model does not see instruments from the test set during training.

The results in \tref{tab:unseen-class} first show that the baseline model conditioned on instrument classes, unsurprisingly, does not perform well on unseen instruments as it is effectively untrained for these classes. 
On the other hand, both few-shot models with $n \in \{1, 5\}$ outperform the baseline by large margins across all instruments (except the 1-shot model for \textit{drums}).
Note that the performance on \textit{drums} is significantly lower than the other two instruments. 
We conjecture that the high variance of drum sounds makes it harder for the model to separate based on just one or few examples at inference time. 
In addition, if we compare \tref{tab:unseen-class} to \tref{tab:onehot-vs-fewshot}, first, we see the drops in SDR resulting from not seeing the testing classes during training. 
Next, we see that the 5-shot model outperforms both the baseline and the 1-shot model more significantly on unseen instruments. 
This aligns with our observations in \sref{sec:baseline-results} 
that the advantage of few-shot MSS is more significant on rare and even unseen instruments. 



 

\begin{table}[t]
\centering
\footnotesize
\begin{tabular}{c cccc}
\toprule
Method & Vocals & Drums & Bass  \\
\midrule
Baseline & $0.35$ & $0.05$ & $-0.10$ \\
\midrule
1-shot & $2.65 \pm 0.74$ & $-0.01 \pm 0.64$ & $1.40 \pm 0.88$ \\
5-shot & \boldmath$3.22 \pm 0.44$ & \boldmath$1.18 \pm 0.40$ & \boldmath$2.88 \pm 0.35$ \\
\bottomrule
\end{tabular}
\caption{Mean SDR (dB) on MUSDB18 for the baseline and few-shot conditioned models with $n \in \{1, 5\}$. Vocals, drums, and bass are held out from the training data. The standard deviations of the means for the few-shot models are computed over 10 test iterations.}
\label{tab:unseen-class}
\end{table}

\subsection{Conditioning examples}
In the previous experiments, we provided few-shot models with single-sourced conditioning examples drawn from the target recording.
To see if few-shot models are able to generalize to more practical scenarios, we evaluate a 5-shot model with relaxed constraints on conditioning examples as discussed in \sref{subsec:conditioning}.

\tref{tab:condition-audio} first shows the performance of the basic 5-shot model where (as before) we apply both constraints to the conditioning examples. 
Then, if we draw conditioning examples from different tracks instead of the target track, SDR drops between $\approx$0.4 to 1.1 dB across instruments. 
Even with this drop, the model still achieves comparable and in some cases better SDRs compared to the baseline model (shown in \tref{tab:onehot-vs-fewshot}). 
On the other hand, if we provide multi-sourced conditioning examples with additional non-target instruments instead of single-sourced ones, the model performance drops significantly. 
This is not surprising given that the model is trained only on isolated audio. 
To see if changing the training conditions would make this scenario possible, we retrain the 5-shot model with multi-sourced conditioning examples.
The retrained model (shown as MS in \tref{tab:condition-audio}), while still performing worse than both the basic 5-shot model and the baseline, achieves about 0.6 to 2.8 dB SDR improvement by matching the training and testing scenarios. 
These results indicate that the proposed few-shot MSS model can adapt to a wide variety of applications where the conditioning examples can be drawn from different recordings. 
The model has a harder time generalizing to multi-sourced conditioning examples, but if that is the desired scenario, we can match the training and inference objectives to achieve better results.

Next, we experiment with including five additional negative conditioning examples, each containing a random non-target instrument, as discussed in \sref{subsec:neg-cond}. 
The results at the bottom of \tref{tab:condition-audio} show that providing additional information about what \textit{not} to separate during training and inference helps to further improve the separation compared to the basic 5-shot model conditioned on positive examples only. This setup could be particularly useful when labeling non-target instruments is much easier than labeling the target one.

\begin{table}[t]
\centering
\footnotesize
\begin{tabular}{P{6em}P{4.8em}P{3em}P{2em}P{2em}P{2.5em}}
\toprule
\multicolumn{2}{c}{Test conditioning constraints} & \multirow{2}{3em}{Model variant} & Vocals & Drums & Bass  \\
\cmidrule(lr){1-2}
Single-sourced & Same track &\\
\midrule
\checkmark & \checkmark & & 4.72 & 4.66 & 3.26 \\
\midrule
\checkmark & & & 4.34 & 4.16 & 2.20  \\
& \checkmark & & 1.00 & 2.03 & -0.63  \\
& \checkmark & MS & 1.63 & 3.52 & 2.21 \\
\midrule
\midrule
\checkmark & \checkmark & +Neg & 4.73 & 5.21 & 3.63 \\
\bottomrule
\end{tabular}
\caption{Mean SDR (dB) on MUSDB18 for the 5-shot model with different constraints applied to the conditioning examples at test time. MS: Model trained with multi-sourced conditioning examples. +Neg: Model trained with both positive and negative single-sourced conditioning examples.}
\label{tab:condition-audio}
\vspace{-1.1mm}
\end{table}

\section{Conclusion}
\label{sec:conclusion}
In this work, we proposed a few-shot MSS paradigm where we condition a generic U-Net source separation model on few audio examples of the target instrument. We evaluated the trained models on real musical recordings in MUSDB18 and MedleyDB datasets. We first quantitatively show that our proposed few-shot MSS model outperformed the baseline model, conditioned on the instrument class labels, on both seen and unseen instruments. Additionally, we saw a more significant advantage of few-shot MSS on rare and unseen instruments.
Next, we further experimented with different characteristics of the conditioning examples, including relaxing the constraints for examples and providing additional negative conditioning. The results indicate the potential of applying few-shot MSS to a wider variety of real-world scenarios, where the conditioning examples may come from different recordings or contain non-target sounds. Future work could explore the structure of the conditioning embedding space and ways to further improve model performance such as different conditioning mechanisms, additional loss terms, and pre-training the conditioning encoder on few-shot classification tasks.

\vfill\pagebreak

\bibliographystyle{IEEEbib}
\bibliography{strings,refs}

\end{document}